\documentstyle[preprint,aps]{revtex}
\tighten
\draft
\begin{document}
\widetext
\preprint{NYU-TH/12/98/01, CLNS 98/1600}
\bigskip
\bigskip
\title{Brane Inflation}
\medskip
\author{Gia Dvali $^{a}$ and  S.-H. Henry Tye $^{b}$}

\bigskip
\address{$^a$ Physics Department, New York University, 4 Washington Place, 
New York, NY 10003 \\ 
and ICTP, Trieste, Italy\\
$^b$ Newman Laboratory of Nuclear Studies, Cornell University, Ithaca, NY 14853}
\date{\today}
\bigskip
\medskip
\maketitle

\begin{abstract}
We present a novel inflationary scenario in theories with low scale (TeV)
quantum gravity, in which the standard model particles are localized on 
the branes whereas gravity propagates in the bulk of large extra dimensions. 
This inflationary scenario is natural in the brane world picture.
In the lowest energy state, a number of branes sit on top of 
each other (or at an orientifold plane), so the vacuum energy cancels out. 
In the cosmological setting, some of the branes "start out" relatively 
displaced in the extra dimensions and the resulting vacuum energy triggers 
the exponential growth of the $3$ non-compact dimensions.
The number of e-foldings can be very large due to the very weak brane-brane 
interaction at large distances. In the effective four-dimensional field 
theory, the brane motion is described by a slowly rolling scalar field 
with an extremely flat plateau potential. When branes approach each other 
to a critical distance, the potential becomes steep and inflation ends 
rapidly. Then the branes "collide" and oscillate about the equilibrium 
point, releasing energy mostly into radiation on the branes.

\end{abstract}
\narrowtext

\subsection{Introduction.}
Recently it was suggested that the fundamental scale of quantum gravity 
$M_{Pf}$ may be as low as TeV, providing an alternative understanding of 
the hierarchy problem\cite{add}. Observed weakness of gravity at large 
distances is associated by $N$ large new dimensions (of size 
$\sim R >> M_{Pf}^{-1}$) in which gravity can propagate. 
Then the relation between the observed Planck scale $M_P$ and the 
fundamental Planck scale $M_{Pf}$ is given by
\begin{equation}
M_P^2 = M_{Pf}^{N + 2}V_N \label{planckscale}
\end{equation}
where $V_N \sim R^N$ ($N$ must be 2 or larger) is the volume in extra 
spatial dimensions ({\em i.e.}, the bulk). In this picture,
all the standard model particles must live in a brane (or a set of branes) 
with $3$ extended space dimensions\cite{add}\footnote{In a different 
context an attempt of lowering the string scale to TeV, without lowering 
the fundamental Planck scale was considered in \cite{lyk}, based on an 
earlier observation in \cite{witten}. Dynamical localization of the fields 
on a (solitonic) brane embedded in a higher dimensional universe has been 
studied earlier in the field theoretic context for spin-$0$ and 
$1/2$\cite{localization1/2}, \cite{localization1} 
and spin-1 states \cite{localization1}}.
Perhaps the most natural embedding of this picture is in the string context
via the $D$-brane construction (see\cite{polchinski} for an introduction), 
where the standard model fields can be identified with open string modes 
stuck on the $D$-branes and the graviton is a closed string mode 
propagating in the bulk \cite{aadd,st,3gen,bw}.

 Various phenomenological and astrophysical constraints\cite{add} and related
issues like black holes\cite{blackholes}, collider physics\cite{collider}, 
flavor violation\cite{flavor}, neutrino masses\cite{neutrino,kakushadze} 
and proton decay\cite{add,kakushadze} have been 
addressed. Unification of the gauge couplings is another important issue 
and in this respect, the proposals of \cite{ddg}, \cite{kakushadze} look 
particularly promising.

 The main concern of the present paper is the inflationary cosmology in this
framework. Obviously, with the above proposal, the standard cosmological 
picture is dramatically affected. As it was shown in \cite{add}, big bang 
nucleosynthesis (BBN) and the bulk graviton production place severe bounds 
both on the reheating temperature and on the "initial conditions" for the 
hot big bang after inflation. First, the reheat temperature must be above
at least $1$ MeV for BBN to proceed successfully. On the other hand too 
much reheating in the bulk can lead to the overproduction of Kaluza-Klein 
(KK) gravitons which can over-close the Universe and ruin the success of BBN. 
Therefore the minimal requirement after inflation would be to
reach a reasonable reheating temperature on the brane $T_R$, while the
bulk remains cold and empty\footnote{For recent discussions on thermal 
phase transitions in higher dimensional theories see \cite{ddgr},\cite{rm}.}.
The question is how to implement inflation in this scenario to achieve 
these properties, in addition to sufficient inflation, and correct 
amounts of density perturbation and baryogenesis? 
Some potential problems in this respect
were considered in \cite{lyth}\cite{kl}. Their considerations, however, were 
limited to conventional inflation potential with the inflaton being one 
of the fields living on the brane. This type of scenarios typically requires 
tremendous fine tuning. 

	In this letter we suggest a novel inflationary scenario, free of the 
above difficulties. The crucial aspect is that, although the inflaton 
is a brane mode in the ground state, it behaves as an "inter-brane" mode 
that describes a relative separation of branes in the extra space directions.
This separation induces inflation in the $3$ non-compact dimensions.
During inflation, when branes are generically separated, the inflaton 
behaves as a weakly coupled field (as it should), with a slow-roll 
behavior. Towards the end of inflation, when branes come close, the 
inflaton couples to the other brane modes but remains 
very weakly coupled to the bulk modes. This means only the brane is reheated,
thus avoiding the problems with late entropy release or over-closure of the
Universe. For the outside observer, this process would look like 
reheating due to the "breathing" of brane-brane bound-state.
An oversimplied inflaton potential that illustrates the above 
generical feature has the form:
\begin{equation}
 V(\phi) \sim M_{Pf}^4 ~\xi(\phi / M_{Pf}) (1 - e^{(- |\phi | /m)})
\end{equation}
where $m$ is a model-dependent mass scale and $\xi(\phi)$ is a
smooth slowly-varying
function.

This inflationary scenario emerges 
rather naturally in the generic brane world picture\cite{add,bw}. We may
consider the 
Type I string where $K$ branes sit more or less on top of an
orientifold
plane at the lowest energy state, resulting in zero cosmological constant.
In the cosmological setting, it is reasonable to assume that some of the 
branes were relatively displaced from the orientifold plane in 
the early universe. This displacement 
induces an effective vacuum energy density triggering inflation. 
The separation distance may 
be identified with the vacuum expectation value $\phi$ of an appropriate 
Higgs field\cite{witten2}. This Higgs field is an open string state with 
its two ends stuck on two separated branes;
that is, this scalar field is a brane mode playing the role of 
the inflaton. In the effective four-dimensional theory, the motion of 
the branes is described by this {\it slowly-rolling} scalar field, 
the inflaton. This has a flat potential with no need of fine tuning.
Flatness of its potential is due to the very weak brane-brane (and 
brane-orientifold plane) force at large distances,
thus allowing for sufficient number of e-foldings. 
When the branes come closer, the short-range potential turns on 
and ends inflation. At this point the inflaton potential becomes very 
steep and its oscillations reheat the brane(s). 
In this scenario, it is even possible that the electroweak Higgs field 
plays the role of the inflaton.
 
\subsection{Brane-induced inflation.}

In this subsection we show how the relative separation of the branes in the 
extra space can lead to the inflation in $3$ non-compact dimensions. 
Consider two $3$-branes\footnote{one can in fact consider them as two 
$p$-branes with $p-3$ dimensions compactified (see the next section).}. 
We will be assuming that there is an attractive potential between these branes
which stabilizes them
 on top of each other.
Let the brane tension, the energy per unit 3-volume, be $T$ and the 
brane-brane separation in the extra space be $r$. 
Then the effective $4$-dimensional vacuum energy density is given by
\begin{equation}
\Lambda_{eff} = V(r) + 2T + \Lambda_{bulk}V_N  \label{lambda}
\end{equation}
where $V(r)$ is the brane-brane interaction energy and $\Lambda_{bulk}$ is an
effective bulk cosmological constant. Their normalization is such that
$V(0) = 0$ and the positive contribution from $T$ is canceled out by 
the negative contribution from $\Lambda_{bulk}$. 
When branes are displaced, $V(r)$ is no longer zero
and results in an effective cosmological constant in four dimensions
$\Lambda_{eff} = V(r)  > 0$. What is the cosmological effect of this term?
Before discussing this, let us make a couple of clarifying remarks. 

To preserve the success of BBN, there is a very strong requirement that 
the size of the extra dimensions should not evolve 
from the time of nucleosynthesis.
On the other hand, for small $N$, it is not desirable to have a large 
reheating temperature in the bulk, since the resulting KK graviton 
excitations can over-close the universe. In the present
context, this is quite natural since the reheating takes place on the
brane(s) only (see below), so we expect little 
change in the bulk temperature in this scenario after inflation. 
Therefore, in the first approximation, we require that extra radii are
frozen both during and after inflation. 
For this to be the case, the radius modulus (radion)
must be stabilized by some effective potential\footnote{We will not try 
to identify an explicit source of the radius stabilization in this paper 
and will take it for granted (see \cite{sundrum},\cite{adm} for possibilities
in this respect and \cite{igncost} for recent discussions on perturbative stability
).}
which can
give it a mass $m_R$ larger than an effective four-dimensional Hubble parameter
\begin{equation}
 m_R > H = (\Lambda_{eff} /3M_P^2)^{{1 \over 2}} \label{radionc}
\end{equation}
This is not a very strong requirement since we expect at most 
$\Lambda_{eff} = V(r) \sim M_{Pf}^4$. 

The second point is that an effective $4$-dimensional Hubble size 
$H^{-1}$ should be larger than the size of extra dimensions $R$. 
Otherwise the universe cannot be treated as four
dimensional at distances $\sim H^{-1}$. 
In view of (\ref{lambda}), this is equivalent to the requirement 
that the ($4 + N$)-dimensional Bulk Hubble length 
\begin{equation}
 H_{bulk}^{-1} \sim
{M_{Pf}^{1 + N/2} \over \Lambda_{bulk}^{1/2}} > R 
\end{equation}
Both of these requirements are easily satisfied for 
$\Lambda_{eff} \sim  T \sim M_{Pf}^4$, even for $N = 2$.

Assuming that the above requirements are satisfied, at distances $ >> R$
the evolution of the $4$-dimensional 
scale factor $a(t)$ is governed by the usual Friedmann
expansion, with an effective Hubble parameter
\begin{equation}
H^2 = \left({\dot a \over a}\right )^2 = {\rho_{eff} \over 3M_P^2}
\label{radionc2}
\end{equation}
where dot denotes a derivative with respect to the cosmic time $t$ and
\begin{equation}
\rho_{eff} = T{\dot r}^2 + V(r)  \label{density1}
\end{equation}
is an effective four-dimensional energy density of the Universe. 
The first term in (\ref{density1}) comes from the kinetic energy of 
relative brane motion (we neglect the center-of-mass motion). 
 Inflation will take place if
\begin{equation}
T{\dot r}^2 << V(r)  \label{infc}
\end{equation}
meaning that $V(r)$ is a flat function for some $r$. 
On the other hand, effective reheating\footnote{By "effective", we 
mean $T_R$ at least up to nucleosynthesis temperature
$T_R \sim 1{\rm MeV}$ on the brane.}
requires $V(r)$ to be steeper at the later stages. 
This is the key point of our
paper: in many cases, $V(r)$ is indeed a steep function for small $r$ rapidly
approaching a constant value for $r \rightarrow \infty$.
To see this, consider the brane-brane interaction at large distances. 
It is governed by the following possible sources:

1) There are modes (both massive and massless) localized on the branes.
Their wave-function decays exponentially fast in the directions
transverse to the brane $\psi(r) \sim e^{-r/r_0}$ where $r_0$ is the 
"thickness" of a brane, an effective localization width of the states. 
Many of the standard model particles
belong to this category.  When branes are separated, the couplings
between the states from different branes are suppressed. 
When the branes are coincident, however, the coupling strengths are 
restored and they contribute to the vacuum energy on the brane. 
{\it Apriori} this contribution cannot vanish since
supersymmetry $must~be$ broken on the brane, where we live. 
Its sign depends on the details.
We assume that this contribution is negative 
(in agreement with the assumed attractive
brane-brane potential) and thus must be canceled out by adjusting a positive
brane tension. This is just the usual fine tuning of the cosmological constant.
On the other hand, when branes are separated, the negative contribution 
decreases
(at least) exponentially fast $\sim - e^{-r/r_0}$ and there must remain
a non-balanced positive constant term at large distances. 
The bottom-line is that, there is a very short range attractive potential
between the branes:
\begin{equation}
V(r) \sim T( 1 - f(r/r_0))
\end{equation}
where $f(r/r_0)$ is an exponentially vanishing function for $r > r_0$.
For solitonic membranes we expect $r_0$ to be at least as small 
as $T^{-{1\over 4}} \sim M_{Pf}^{-1}$.

2) Another potential source is an exchange of the massive bulk modes, which at
large distances ($r >> m^{-1}$) decouple and create an 
Yukawa suppressed potential $\sim  r^{2-N} {e^{-mr}}$.

3) In the same way, an exchange of the light bulk modes, like graviton or 
gauge fields will generate a power-law potential $\sim 1/r^{N -2}$; 
an over-all sign depends on the balance between the tension and 
gauge charge density on the brane\footnote{In some cases,
log$r$ dependence should be understood for the $N = 2$ case.}. 
For instance, gravitational interaction is attractive between $D$-branes 
and repulsive between $D$-branes and the orientifold plane, while the 
R-R fields have an opposite effect
and in the unbroken supersymmetry limit can compensate
the graviton/dilaton force. In this limit, branes are
BPS states. However, even if the RR fields and the dilaton field get 
small masses, their effect
can still "screen" gravity at distances $r << m^{-1}$.

4) Confining potential $\sim kr$ due to the strings stretching 
between the branes. One expects open strings stretched between branes. 
This is true both for $D$-branes\cite{polchinski}
as well as for field theoretic solitons\cite{localization1}. In the latter case
they are flux tubes which are presented due to the fact that bulk must be
confining (in order to have a massless gauge field localized on the brane)
\cite{localization1} and therefore the
gauge flux is trapped in the flux tubes that can stretch from brane to brane.
Thus, at large distances one expects the brane-brane potential to be
\begin{equation}
V(r) = T(\alpha - f(r/r_0) + b_i{e^{-m_ir} \over r^{N - 2}} + 
{c\over r^{N -2}} + kr)
\label{formofpotential}
\end{equation}
where $\alpha, b_i, c$ are model-dependent constants.
$k$ is proportional to the density of the stretched strings per unit
$3$-volume. During inflation, $k$ is redshifted away exponentially
fast $k(t) \sim k_{in} e^{-3Ht}$. 
The condition that the linear term will allow inflation is
\begin{equation}
{M_Pk_{in}\over M_{Pf}^2 (aT + k_{in}r_{in})} < 1
\end{equation}
at least in some region of $H^{-1}$ size. 
The corresponding region will then expand making $k$ negligible very 
rapidly\footnote{Note that for the regions with $k\sim M_{Pf}^{-1}$
inflation would require either $r_{in}\sim M_P/M_{Pf}^2 \sim 1$mm, or 
$T < M_{Pf}$.}.
In what follows we will assume that this is the case and ignore the 
linear term.

We now turn to the discussion of the resulting inflationary dynamics. It
is most convenient to do this in the effective  $4$-dimensional field theory
picture, where brane-brane separation is described
by an expectation value of the real scalar field (inflaton):
\begin{equation}
\phi \sim M_{Pl}^2r  \label{phir}
\end{equation}
There is a simple understanding of this relation 
(see the next section for an alternative
explanation in the $D$-brane context) 
if one
thinks of branes as some sort of the field theory solitons
formed by the massive scalar
fields $\Phi_1$ and $\Phi_2$ with the Lagrangian
\begin{equation}
L = L_1(\Phi_1) + L_2(\Phi_2) + L_{cross}(\Phi_1\Phi_2)
\end{equation}
In this language branes are stable configurations
independent of $x_{\mu}$ and localized in the external space at
$x_A = 0$ and $x_A = r_A$ respectively
\begin{equation}
\Phi_1 = \Phi_1(x_A),~~~~\Phi_2 = \Phi_2(x_A - r_A)
\end{equation}
(where the index $A = 1,2..,N$ labels the extra coordinates).
It is most crucial that away from the branes the situation
looks like the translational invariant vacuum
up to an {\it exponentially small} correction $\sim e^{-m|x|}$. Assume for a
moment that there is no cross interaction among $\Phi_1$ and $\Phi_2$
fields in the Lagrangian ($L_{cross} = 0$). Then there are $N$
massless modes localized on
{\it each}
brane, corresponding to $N$ transverse excitations:
\begin{equation}
\Phi_i(x_A + G^i_A(x_{\mu})) = {\rm brane} + G^i(x_{\mu})
{\partial\Phi_i(x_A) \over \partial x_A} +  {\rm heavy~modes}
\end{equation}
These are Goldstone modes of spontaneously broken translational invariance.
However, in the presence of two branes only $N$ superpositions ($G_A^1 +
G_A^2$), corresponding to the center of mass motion,
are true Goldstones modes\footnote{In the presence of gravity, these are 
eaten up by graviphotons $g_{\mu A}$ and get mass 
$\sim T^{{1 \over 4}} M_{Pf}/M_P$ \cite{add}}.
The remaining $N$ orthogonal combinations
$\phi_A(x_{\mu}) = G_A^1 - G_A^2$ correspond to relative vibrations
of the branes and are massless only because membranes are non-interacting.
However, even in the presence of local cross couplings, e.g.
\begin{equation}
L_{cross} = \lambda(\Phi_1\Phi_2)^2
\end{equation}
the induced potential is {\it short range}, due to localized nature of branes,
and decays exponentially fast for large separation.
Corresponding four-dimensional mode describing this separation is given by
\begin{equation}
\phi = \left ( (r_A + G_A^1 - G_A^2)^2 \right) ^{{1 \over 2}}
\end{equation}
and has an exponentially decaying potential for large field values
\begin{equation}
V(\phi) \sim  T(\alpha - f(\phi/M))
\end{equation}
where $f(\phi/M)$ goes at least as $\sim  e^{-\phi/M}$ and 
$M \sim T^{{1 \over 4}}$ is a typical mass parameter of the theory.
As discussed above, the constant term $\alpha$ comes from the short 
range binding energy of the solitons, which at the tree level is roughly:
\begin{equation}
E_{binding} \sim \int dV_N \left ( L_{cross}(\Phi_1(x_A)\Phi_2(x_A)) -
L_{cross}(\Phi_1(x_A)\Phi_2(\infty)) \right ) \label{binding}
\end{equation}
There will be an additional contribution to $\alpha$ and $f$ from the modes 
localized on the solitons. All this contributions are very short 
range bounded by the brane thickness.
In the presence of light bulk modes, such as gravity or gauge fields, 
there will be an additional
interaction of the form (\ref{formofpotential}) so that the effective potential
will become:
\begin{equation}
V(\phi) \sim  T \left(\alpha - f(\phi/M) +  {1 \over |\phi|^{N - 2}} \left (\beta_i
e^{-|\phi|/M_i} + \gamma\right )\right)
\label{pot}
\end{equation}
where index $i$ runs over all the massive modes. 
Note that $M_i \sim M_{Pf}^2/m_i$ depend on the masses
of bulk modes and {\it apriori} may be
totally unrelated to the value of $T$.

Let us now turn to the inflationary dynamics following from the above 
potential.
We will assume that all the bulk masses are $m_i < M_{Pf}$ and, therefore,
for $\phi >> m_i^{-1}$ the short range potential $f(\phi/M)$ plays no
role. Note that for $m_i << R^{-1}$ case, it is possible that $\beta_i$ and $c$
terms conspire and cancel out for $r << m^{-1}_i$. 
In such a case $V(r)$ can be dominated
by $f(r)$ term. This may happen in the string theory and will be discussed 
later. For the moment however we assume that there is no such a conspiracy 
and will discuss the case of very small $\gamma$ first.

As mentioned above, we assume that Universe starts out in the state
with nonzero vacuum energy induced
by (parallel) branes separated at some distance $r_{in} \sim  \phi_{in}$
and slowly approaching each other under the action of an attractive force.
In the effective four-dimensional language this is described by the
slowly-rolling scalar field $\phi$, whose evolution is governed
by the following equation of motion
\begin{equation}
{d^2\phi \over dt^2} + 3H{d\phi \over dt} + {dV \over d\phi} =0 \label{rolling!}
\end{equation}
where $V$ is given by (\ref{pot}) and we take $\gamma = 0$ to start with.
Standard slow-roll conditions are
\begin{equation}
|M_P V'/V| << 1,~~~|M_P^2V''/V| << 1 \label{slow-rollc}
\end{equation}
where prime denotes the derivative with respect to $\phi$.
Breakdown of either of these will mark the end of inflation. The corresponding
value $\phi_{end}$ is determined by:
\begin{equation}
{\beta \over \alpha \phi_{end}^{N-2}}e^{-{\phi_{end} \over M}} \simeq \left
({M\over
M_P }\right )^2  \label{breakdown!}
\end{equation}
The number of $e$-foldings $n$ between two values $\phi_{end}$ and
$\phi_n$ can be found by integrating (\ref{rolling!}) subject to the
slow roll (basically ignoring the second derivative)
\begin{equation}
n = {1 \over M_P^2}\int_{\phi_{end}}^{\phi_n}d\phi {V \over dV/d\phi}
\end{equation}
This gives
\begin{equation}
{\beta\over \alpha \phi_n^{N-2}}e^{-{\phi_n \over M}} \simeq \left ({M\over
M_P }\right )^2 {1 \over (n + 1)} \label{nbeforetheend}
\end{equation}
Comparing with (\ref{breakdown!}) we see that during the last $60$ e-folds,
the change in $\phi$ is very small. Thus for small $\gamma$ the requirement 
of successful inflation puts basically no
constraint on the number and size of extra dimensions as far as
$R >> m^{-1}_i$. For $\gamma \sim \beta_i$ the slope of the potential at $\phi
>> M_i$ is dominated by the power-law term and the slow-roll breaks down at
\begin{equation}
\alpha\phi_{end}^N = \gamma M_P^2(N-1)(N-2)
\end{equation}
and the value $\phi_n$ ($n$ e-folds before the end of inflation) is given by
\begin{equation}
\alpha \phi_n^N = (n + 1)\gamma M_P^2 N(N-2) \label{efolds}
\end{equation}
Now using equations (\ref{planckscale}) and (\ref{phir}) and taking into 
account that the
maximal initial separation of the branes is bounded by the size of extra
dimension $r_{in} < R$, we get that for having $60$ e-foldings with
maximal initial separation, $c$ must satisfy
\begin{equation}
 M_{Pf}^{N-2} > 60 (N-1)(N-2)\gamma/\alpha
\end{equation}
which is not a significant constraint.

\subsection{The Type I String Picture}

        The observed universe contains gauge and matter fields as
well as gravity. Since superstring theory\cite{polchinski} is the
only known theory that incorporates consistent quantum gravity, one
would like to see how string theory can describe our universe.
In the brane world picture\cite{add,aadd,bw}, gravity lives in the 
ten dimensional space-time,
whereas gauge fields can be localized on $p$ spatial dimensional extended
objects known as $Dp$-branes. In particular, our four dimensional world
(including the strong and electroweak interactions as well as the quarks and
leptons) resides inside a set of overlapping branes (or intersection
thereof with other branes), with the extra $p-3\geq 0$
spatial dimensions compactified on a manifold with some finite volume
$V_{p-3}$. However, gravity is free to propagate in the
ten dimensional bulk of the space-time with the remaining $9-p$ spatial
dimensions compactified on a manifold with some finite volume $V_{9-p}$.
Dilaton (and other moduli) stabilization generically requires the string
coupling to be $g_s
{\ \lower-1.2pt\vbox{\hbox{\rlap{$>$}\lower5pt\vbox{\hbox{$\sim$}}}}\ } 1$.
Thus, in the brane world picture, the four dimensional gauge and
gravitational couplings
scale as $1/V_{p-3}$ and $1/V_{p-3} V_{9-p}$, respectively.
By tuning $V_{9-p}$ (for $p<9$) we can achieve gauge and gravitational
coupling unification.
Then for $p>3$, we can choose $V_{p-3}$ large enough so
that the four dimensional gauge couplings are small even if
$g_s$ is not. The $3<p<9$ feature means that our world is inside
a $p$-brane. In this brane world picture, the string scale $M_s$ can
{\em apriori} be anywhere between the electroweak scale and the
Planck scale. Physics becomes most exciting when $M_s$ is around
TeVs\cite{add}. Note that
\begin{equation}
M_P^2 \sim M_s^8 V_{p-3} V_{9-p}
\end{equation}
Comparing with (\ref{planckscale}), we obtain the relation
\begin{equation}
M_{Pf}^{2+N} \sim M_s^8 V_{p-3} V_{9-p-N} 
\end{equation}
Here $M_{Pf}$ is expected to be somewhat larger than $M_s$, but within 
the same order of magnitude.

        To see that the inflaton potential during the inflationary epoch
can appear rather naturally in the brane world, let us consider the 
embedding of the TeV scale scenario in
the perturbative 4-dimensional Type I string \cite{aadd,type1}, where some
semi-realistic models have been constructed \cite{st,3gen}. 
A typical model will have orientifold planes with negative tensions 
(i.e., energy density)\cite{Dai}. It turns out that the orientifold 
plane is also charged under some ($p+1$ form) tensor (i.e., RR) fields 
present in the theory. One may view these planes as rigid $p$-branes 
with negative brane tensions and negative RR charges.
(To be concrete, one may choose $p=5$.)
When all the $9-p$ directions are compactified, the negative RR
charges of the orientifold plane must be canceled by the introduction of
positively charged D-branes. In the classical vacuum of a Type I string,
a set of D-branes sit on top of each orientifold plane, with zero net RR
charge. Let the brane tension of a D$p$-brane be $T_p$. Then the sum of the
energy density $KT_p$ of the set of $K$ D$p$-branes exactly cancels the
negative energy density -$KT_p$ of the orientifold plane ($K=2^l$). 
In a supersymmetric vacuum,
the energy density is exactly zero everywhere. (There are vacua in which the
branes are separated. F theory\cite{vafa} provides the proper description of
these generalized situations.)

        In the early universe, however, the D$p$-branes do not have to be
exactly on top of the orientifold plane. Let us consider the simple 
situation (this is also the most relevant situation,
as we shall explain later) where a single D$p$-branes is separated
from the rest by a distance $r$.
The tension (energy density) of the brane, $T_p \sim M_s^{p+1}$, is
opposite to the sum of the enrgy density of the remaining branes 
plus the orientifold plane.
Before dilaton (and other moduli) stabilization and SUSY breaking, the
static potential due to the exchange of the closed string sector fields 
has the form \cite{bachas}
\begin{equation}
V(r) \sim T_p (1-1) F(r) 
\end{equation}
where $r$ is the displacement distance in the extra large directions, 
and the function $F(r)$ is model dependent. Crudely speaking,
$F(r)$ is the dual of the one-loop open string amplitude.
This potential arises from the exchange of closed string (bulk) states
between the displaced D$p$-brane and the rest. The closed string spectrum 
can be separated into two groups, the NS-NS states and the RR states.
The positive term comes from the exchange of NS-NS
states and the negative terms come from the exchange of the RR states.
The massless NS-NS states
include the graviton and the dilaton-axion. The force due to their
exchanges is repulsive. The exact cancellation of the forces due
the NS-NS and RR fields is a consequence of the BPS property of the branes.
If the brane is moving slowly towards the rest, velocity-dependent terms
are expected in the potential. Before supersymmetry breaking, the leading 
term has the form $v^4 r^{p-7}$. After supersymmetry breaking, we expect 
a $v^2$ term to be present.

        In the more realistic situation, where the string model describes
our universe, the dilaton (and other moduli) must be stabilized dynamically,
and supersymmtry must be broken (presumably softly and dynamically).
To be compatible with experiments, all the bosonic bulk modes except 
the graviton are expected to become massive. This means that the 
dilaton-axion and the RR fields must become massive (so that their 
long range forces are Yukawa suppressed), while only gravity remains 
long ranged.
After the stabilization of the dilaton and other moduli,
we expect the potential $V(r)$ to be, at large $r$,
\begin{equation}
V(r)=M_{Pf}^{6-N}~ r^{2-N}(1 + \sum e^{(-m_i r)} - \sum e^{(-m_j' r)}) 
\label{stringy}
\end{equation}
where $m_i$ are the masses of the NS-NS states while $m_j'$ are the masses
of the RR fields. For large $r$ and N$=2$, $V(r)$ is essentially a constant.
For $r=0$, we expect $V(r=0)=0$, since today's cosmological constant is 
(almost) zero. For small $r$, the form of $V(r)$ depends crucially on the 
mass spectrum. However, the overall shape of $V(r)$ is quite clear: it 
rises rapidly from zero at $r=0$ to the $r^{2-N}$ form for large $r$.

Besides the massive states ({\em i.e.}, $\sim M_{Pf}$ or larger) which we may 
safely ignore, there are bulk light states which were massless before 
supersymmetry breaking and dilaton stabilization. 
For experimental compatibility, it is enough that their masses
are of the order of inverse millimeter(mm) or larger. 
(From the experimental view-point, this is a very exciting scenario, due 
to the sub-millimeter test of Newton's gravitational law\cite{test}).
This will automatically be the case if the only stabilizing potential for 
them comes from supersymmetry breaking on the brane\cite{aadd}. In this case, of
course,
their contributions will keep canceling gravitation at distances smaller than
inverse mm. Effectively, the coefficient of $1/r^{N-2}$ term 
($c$ in the previous section) will be suppressed at least by 
a factor $ \sim {M_{Pf}^2 \over M_P}$, since this measures the 
transmission of the supersymmetry breaking in the brane to the bulk
and thus generates the dilaton-axion-RR masses.
So the number of e-foldings can be very large! 
In general, the possible number of e-foldings
will be bounded by Eq.(\ref{efolds}).

In the TeV scenario, supersymmetry must be broken on the brane in 
full strength, so we expect the appearance of the constant term in 
(\ref{stringy}), of the form $T\alpha$, which should come from 
the short-range brane-brane binding energy, an analog of
(\ref{binding}), as well as Higgs potential from the brane modes. 

Now we want to argue that the separation distance $r$ is really a 
brane mode, not a bulk mode. This is crucial to the reheating problem.
Suppose the displaced brane belongs to a set of branes that yield
a $SU(L)$ gauge symmetry, {\em i.e.}, the $SU(L)$ gauge bosons are open 
strings whose ends live on this set of branes. Then the 
separation of the branes breaks the $SU(L)$ gauge symmetry, resulting in
some massive gauge bosons\cite{witten2}. 
The energy due to
a stretched open string is $M_s^2r$ (i.e., string tension times length).
However, we know that this corresponds to the mass of a gauge boson after
spontaneous symmetry breaking,
\begin{equation}
 mass = g \phi \sim M_s^2 r
\end{equation}
so we can identify $r$ with the vev of the Higgs field $\phi$. Here $\phi$
is an open string state whose one end is stuck on the displaced brane, 
while its other end is stuck on one (or an appropriate linear 
combination) of branes that are sitting on the orientifold plane. 
So it is a brane mode. (At the ground state,
$\phi$ lives mostly inside the branes.) After the integration 
over the extra large dimensions, the above 
potential $V(r)$ becomes the potential term $V(\phi)$ in 
the 4-dimensional effective field theory, so $\phi$ becomes the inflaton.
More generally, even when there is no gauge symmetry associated with 
the separation of a particular brane, the separation distance $r$ 
is identified with the vacuum expectation value of an appropriate scalar
field; that is, this scalar field is a brane mode playing the role of
the inflaton. 
In the effective four-dimensional theory, the motion of
the branes is described by this {\it slowly-rolling} scalar field,
the inflaton. This has a flat potential with no need of fine tuning.

In the early universe, some number of open strings are stretched between
the branes. The density of such stretched strings (or
other stretched branes) depends on the initial condition/situation.
This density will decrease as the universe expands. In fact, inflation 
will shift it to zero rapidly.

At $\phi=0$, we know $V(\phi=0)=0$, so that the cosmological constant 
remains zero. At small $\phi$, we know the potential is quite steep.
In fact, the precise shape is very model/dynamics dependent. Fortunately,
the details there are not so important. Since $\phi$ is a brane mode, its
coupling to closed string ({\em i.e.}, bulk) modes are extremely weak, 
while its coupling to other open string ({\em i.e.}, brane) modes are
around typical gauge coupling strength, so its damping reheats predominantly the 
brane but not the bulk.

In this brane inflation scenario, the inflaton could have been the
electroweak Higgs field in the standard model. In this case,
the minimum of the potential should be shifted away from the $\phi=0$ point
to the electroweak scale. This can be achieved if the electroweak Higgs
potential is generated after supersymmetry breaking.

Now we can consider the general case, where the branes are all separated 
from the orientifold plane. The separations may even be at different angles.
A careful treatment requires an analysis in F theory\cite{vafa}. However, 
we can easily envision that, one after another,
the branes begin falling on top of the orientifold, until only one brane,
or a set of overlapping branes, remains separated from the rest. As long
as this separation survives 60 e-foldings, the details before that is
irrelevant to observations today. 

\subsection{Density Perturbation and Baryogenesis.}

Now let us discuss the issue of density perturbations in our scenario. 
The predicted value of ${\delta \rho \over \rho}$ can be estimated 
from the standard formula
\begin{equation}
{\delta \rho \over \rho} \sim {H \over M_P\epsilon}
\end{equation}
where $\epsilon = M_P V'/V$ is the slow-roll parameter in (\ref{slow-rollc}).
The right-hand side of the above equation must be evaluated at a time
when the scale of interest crosses out of the de Sitter horizon $H^{-1}$.
Let the corresponding inflaton value be $\phi_n$. 
Perturbations on the present Hubble scales
correspond to the value of ${\delta \rho \over \rho}$ at $\phi = \phi_{60}$.
In the usual picture (with $M_{Pf} \sim M_P$) the Hubble parameter can be 
reasonably large ($\sim 10^{13} - 10^{14}$GeV or so) and therefore the 
observed value of ${\delta \rho \over \rho} \sim 10^{-5}$ does not
require very small $\epsilon_{60}$. In our case the Hubble parameter is
tiny ${H \over M_P} \sim 10^{-32}$ and thus $\epsilon_{60}$ must be 
enormously small ($10^{-28}$ or so!). This means that the potential 
at $\phi_{60}$ must be extremely flat and sharply become very steep at 
$\phi_{end}$. This can be achieved if the inflaton potential is 
dominated by the short range brane-brane interaction. 
Very crudely,
the density perturbations then come out to be
\begin{equation}
{\delta \rho \over \rho} \sim {1 \over r_{60} M_P} 
\label{density}
\end{equation}
We see that (in agreement with \cite{kl}) the requirement
of large density perturbations places more severe constraint on the
flatness of the inflaton potential than the requirement of successful
inflation with $60$ e-foldings, which in our scenario requires no
fine-tunning. An expected lower limit for $r_{60}$ is around $M_{Pf}^{-1}$,
which for $M_{Pf} \sim$ TeV would give ${\delta \rho \over \rho} \sim
10^{-15}$. The observed value would require branes with a very short
range potential 
($\sim  10^4M_P^{-1}$). This would mean that a brane cannot 
fluctuate at scales $>> 10^{-28}$cm and is an extremely rigid object.

There can be other sources of the density fluctuations, e.g.
coming from the branes falling on the
orientifold plane earlier, $60$ e-folds before the end of inflation
(see discussion at the end of the previous section). To conclude,
the precise origin of the density perturbations in the present context is
an open question and requires an additional study.

Another important issue in theories with TeV scale quantum gravity
is baryogenesis. As mentioned above, our scenario gives a natural posibility
of releasing the inflaton energy predominantly into the branes.
This is a desirable starting point for avoiding the bulk graviton 
over-production.
However, even in such a optimistic case the reheating temperature on the 
brane cannot be arbitrary high since the bulk gravitons can be produced
by {\it brane-evaporation}\cite{add}. In particular, reheating up to the 
electroweak temperatures is problematic (at least for small $N$) and thus 
the standard scenarios of baryogenesis is not operative. 
However, in our case, there are new possible sources:
first, since the Hubble parameter is so small, the electroweak 
transition may in fact coincide with the inflationary one and be 
{\it non-thermal}. This will provide an out-of-equilibrium condition. 
More importantly, the inflaton can directly decay into the baryons via 
the baryon-number- and CP-violating $M_{Pf}$-suppressed interactions.
Such operators can be harmless for proton decay due to unbroken 
discrete subgroups of $B$ and $L$\cite{aadd} (see \cite{kakushadze} 
for an explicit construction). Thus, in principle at least, the baryon 
number can be generated with a very low reheating temperature.

\subsection{Conclusions}
 
In this paper, we present a novel brane inflationary scenario in theories
with low scale (TeV) quantum gravity.
This inflationary scenario is natural in the brane world picture.
In the classical vacuum, $K$ branes, with positive energy densities,
sit on top of an orientifold plane, with negative energy density, so
the vacuum energy cancels out. In the early universe, some of the
branes "start out" relatively displaced in the extra dimensions.
This introduces a vacuum energy, which triggers inflation in the 
non-compact 3-dimensional space.
The number of e-foldings can be very large due to the very weak 
brane-brane interaction at large distances, which is mainly due to 
the graviton and other massive
bulk mode exchanges. The displacement distances between branes
are identified with the vacuum expectation values of scalar fields, which
are brane (open string) modes. When branes are close to each other,
the potential
becomes steep and inflation ends rapidly, releasing energy mostly into
radiation on the branes. In this scenario, the electroweak Higgs
field may even play the role of the inflaton.

Although this brane inflationary scenario emerges rather naturally in the
brane world picture and the overall picture looks promising, its details
are quite model-dependent. Precise predictions seem to require a better
understanding of the moduli stabilization and supersymmetry breaking
mechanism. On the other hand, we can use the inflationary requirements to
further constraint the brane world dynamics. For example, the shape of the
inflaton potential depends strongly on the masses of the light bulk
(closed string) modes, which are massless before moduli stabilization
and supersymmetry breaking. They are expected to have tiny masses if
supersymmetry breaking occurs on the brane and is then transmitted to
the bulk.

Although the brane inflation can easily reheat the brane to a 
temperatures around the electroweak scale $M_{ew}$, for small number 
of extra dimensions and low $M_{Pf}$, this is forbidden, due to the 
subsequent evaporation into the bulk gravitons\cite{add}. Thus the 
standard electroweak baryogenesis cannot be applied in such a case. 
However, in the brane inflation context, there are new potential
sources, which can allow baryon number generation at much lower 
temperatures. This is due to the possibility of direct $B$- and 
$CP$-violating decays of the inflaton into the brane-baryons.
It is important to study this issue in greater detail.

\acknowledgments

{}We thank Philip Argyres, Gregory Gabadadze, Zurab Kakushadze, 
Gary Shiu and Alex Vilenkin for discussions. 
The research of S.-H.H. Tye is partially supported by 
the National Science Foundation.


\begin{references} 

\bibitem{add} N. Arkani-Hamed, S. Dimopoulos and G. Dvali, 
Phys. Lett. {\bf B429} (1998) 263, hep-ph/9803315; hep-ph/9807344, 
Phys. Rev. {\bf D} (to be published).

\bibitem{lyk} J. Lykken, Phys. Rev. {\bf D54} (1996) 3693.

\bibitem{witten} E. Witten, Nucl. Phys. {\bf B471} (1996) 135.

\bibitem{localization1/2} V.A. Rubakov and M.E. Shaposhnikov, Phys. Lett. 
{\bf B125} (1983) 136; see also, \\
A. Barnaveli and O. Kancheli, Sov. J. Nucl. Phys. {\bf 51} (1990) 573.

\bibitem{localization1} G. Dvali and M. Shifman, Nucl. Phys. {\bf B504} 
(1997) 127; Phys. Lett. {\bf B396} (1997) 64.

\bibitem{polchinski} J. Polchinski, "String Theory", Cambridge University 
Press, 1998.

\bibitem{aadd} I. Antoniadis, N. Arkani-Hamed, S. Dimopoulos and G. Dvali, 
Phys. Lett. {\bf B436} (1998) 257, hep-ph/9804398.

\bibitem{st} G. Shiu and S.-H.H. Tye, Phys. Rev. {\bf D58} (1998) 106007, 
hep-th/9805147.

\bibitem{3gen} Z. Kakushadze, Phys. Lett. {\bf B434} (1998) 269;
Nucl. Phys. {\bf B535} (1998) 311; Phys. Rev. {\bf D58} (1998) 101901;\\
Z. Kakushadze and S.-H.H. Tye, Phys. Rev. {\bf D58} (1998) 126001.

\bibitem{bw} Z. Kakushadze and S.-H.H. Tye, hep-th/9809147.

\bibitem{blackholes} P.C. Argyres, S. Dimopoulos and J. March-Russell, 
hep-th/9808138.

\bibitem{collider} 
G.F. Giudice, R. Rattazzi and J.D. Wells,  hep-ph/9811291; \\
S. Nussinov and R. Shrock, hep-ph/9811323; \\
E.A. Mirabelli, M. Perelstein and M.E. Peskin,  hep-ph/9811337; \\
T. Han, J.D. Lykken and R.-J. Zhang, hep-ph/9811350; \\
J.L. Hewett, hep-ph/9811356.

\bibitem{flavor} N. Arkani-Hamed and S. Dimopoulos, hep-ph/9811353; \\
Z. Berezhiani and G. Dvali, hep-ph/9811378.

\bibitem{neutrino} N. Arkani-Hamed, S. Dimopoulos, G. Dvali and J.
March-Russell, hep-ph/9811448; \\
K.R. Dienes, E. Dudas and T. Gherghetta, hep-ph/9811428.

\bibitem{kakushadze} Z. Kakushadze, hep-ph/9811193; hep-th/9812163.

\bibitem{ddg} K.R. Dienes, E. Dudas and T. Gherghetta, Phys. Lett.
{\bf B436} (1998) 55, hep-ph/9803466; hep-ph/9806292; hep-ph/9807522.
 
\bibitem{lyth} D. Lyth, hep-ph/9810320.

\bibitem{kl} N. Kaloper and A. Linde, hep-th/9811141.

\bibitem{ddgr} K.R. Dienes, E. Dudas, T. Gherghetta and A. Riotto, 
hep-ph/9809406. 

\bibitem{rm} A. Riotto and M. Magiore, hep-ph/9811089.


\bibitem{witten2} E. Witten, Nucl. Phys. {\bf B443} (1995) 85.

\bibitem{sundrum} R. Sundrum, hep-ph/9805471; hep-ph/9807348.

\bibitem{adm} N. Arkani-Hamed, S. Dimopoulos and J. March-Russell, 
hep-th/9809124.
\bibitem{igncost} I. Antoniadis and C. Bachas, hep-th/9812093.
\bibitem{type1} For a partial list, see, {\em e.g.},\\
G. Pradisi and A. Sagnotti, Phys. Lett. {\bf B216} (1989) 59;\\
M. Bianchi and A. Sagnotti, Phys. Lett. {\bf B247} (1990) 517; Nucl. Phys.
{\bf B361} (1991) 539; \\
E.G. Gimon and J. Polchinski, Phys. Rev. {\bf D54} (1996) 1667; \\
A. Dabholkar and J. Park, Nucl. Phys. {\bf B472} (1996) 207,
hep-th/9602030; Nucl. Phys. {\bf B477} (1996) 701, hep-th/9604178;\\
E.G. Gimon and C.V. Johnson, Nucl. Phys. {\bf B477} (1996) 715,
hep-th/9604129; \\
J. Polchinski, Phys. Rev. {\bf D55} (1997) 6423, hep-th/9606165; \\
M. Berkooz and R.G. Leigh, Nucl. Phys. {\bf B483} (1997) 187;\\
C. Angelantonj, M. Bianchi, G. Pradisi, A. Sagnotti and
Ya.S. Stanev, Phys. Lett. {\bf B385} (1996) 96;\\
Z. Kakushadze, Nucl. Phys. {\bf B512} (1998) 221;\\
Z. Kakushadze and G. Shiu, Phys. Rev. {\bf D56} (1997) 3686; Nucl.
Phys. {\bf B520} (1998) 75;\\
J.D. Blum and A. Zaffaroni, Phys. Lett. {\bf B387} (1996) 71; \\
G. Zwart, Nucl. Phys. {\bf B526} (1998) 378;\\
G. Aldazabal, A. Font, L.E. Ib{\'a}{\~n}ez and G. Violero, hep-th/9804026;\\
Z. Kakushadze, G. Shiu and S.-H.H. Tye, Nucl. Phys. {\bf B533} (1998) 25;\\
R. Blumenhagen and A. Wisskirchen, Phys. Lett. {\bf B438} (1998) 52;\\
L.E. Ib{\'a}{\~n}ez, R. Rabadan and A.M. Uranga, hep-th/9808139.

\bibitem{Dai} J. Dai, R.G. Leigh and J. Polchinski, Mod. Phys. Lett.
{\bf A4} (1989) 2073; \\
R.G. Leigh, Mod. Phys. Lett. {\bf A4} (1989) 2767; \\
P. Hovara, Nucl. Phys. {\bf B327} (1989) 461; 
J. Polchinski, Phys. Rev. Lett. {\bf 75} (1995) 4724.

\bibitem{vafa} C. Vafa, Nucl. Phys. {\bf B469} (1996) 403; \\
D. Morrison and C. Vafa, Nucl. Phys. {\bf B473} (1996) 74;
Nucl. Phys. {\bf B476} (1996) 437.

\bibitem{bachas} For a review, see {\em e.g.}, 
C.P. Bachas, hep-th/9806199.
\bibitem{test} J.C.Price, in proc. Int. Symp. on Experimental Gravitational Physics,
ed. P.F. Michelson, Guangzhou, China (World Scientific, Singapore 1988); J.C. Price
et. al., NSF proposal 1996; A. Kapitulnik and T. Kenny, NSF proposal, 1997; J.C. Long,
H.W. Chan and J.C. Price, hep-ph/9805217.

\end{references}
\end{document}